\documentclass[conference]{IEEEtran}

\usepackage{amsmath}
\usepackage[T1]{fontenc}
\usepackage[utf8]{inputenc}
\usepackage[font=small,labelfont=bf,tableposition=top]{caption}
\usepackage{blindtext}
\usepackage{cuted}
\usepackage{array}

\usepackage{xcolor}
\usepackage{color,balance}
\usepackage{subfiles,paralist}
\usepackage{enumitem}
\usepackage{mathtools}

\usepackage{amsmath}
\usepackage{amssymb}
\usepackage{amsthm}
\usepackage{listings}
\usepackage{bbm}
\definecolor{dkgreen}{rgb}{0,0.6,0}
\definecolor{gray}{rgb}{0.5,0.5,0.5}
\definecolor{mauve}{rgb}{0.58,0,0.82}

\usepackage{thmtools, thm-restate}
\usepackage{mathtools,extarrows}
\theoremstyle{definition}
\newtheorem{definition}{Definition}[section]

\theoremstyle{remark}
\newtheorem*{remark}{Remark}
\usepackage{hyperref}
\usepackage[capitalise, noabbrev]{cleveref}
\usepackage{subcaption}
\usepackage{graphicx}
\usepackage{adjustbox}
\usepackage{tikz}
\usepackage[boxruled,linesnumbered, ruled,vlined]{algorithm2e}
\usepackage[noend]{algpseudocode}
\usepackage[most]{tcolorbox}

\newcommand{\RNum}[1]{\uppercase\expandafter{\romannumeral #1\relax}}

\usepackage{xpatch}
\xapptocmd\normalsize{%
 \abovedisplayskip=4pt
 \abovedisplayshortskip=0pt plus 4pt
 \belowdisplayskip=4pt
 \belowdisplayshortskip=5pt plus 3pt minus 4pt
}{}{}
\usepackage{diagbox}
\usepackage[normalem]{ulem}
\usepackage{multirow}
\usepackage{colortbl}

\newcommand{\mempool}{$\mathbb{T}$}
\newtheorem{corollary}{Corollary}

\renewcommand\paragraph[1]{\vspace{0.3mm}\noindent \textbf{#1.\ }}
\newcommand\numberthis{\addtocounter{equation}{1}\tag{\theequation}}

\DeclareMathOperator*{\argmax}{arg\,max} 

\begin{document}


\title{Towards Overcoming the Undercutting Problem} 
\author{
\IEEEauthorblockN{
    Tiantian Gong\IEEEauthorrefmark{2},
    Mohsen Minaei\thanks{\IEEEauthorrefmark{1} Part of this work was done while the author was at Purdue University.}\IEEEauthorrefmark{3}\IEEEauthorrefmark{1},
    Wenhai Sun\IEEEauthorrefmark{2},
    Aniket Kate\IEEEauthorrefmark{2}
}
\IEEEauthorblockA{\IEEEauthorrefmark{2} Purdue University,\quad Email: \{tg, sun841, aniket\}@purdue.edu}
\IEEEauthorblockA{\IEEEauthorrefmark{3} Visa Research,\quad Email: mominaei@visa.com}

}

\IEEEoverridecommandlockouts

\maketitle

\begin{abstract}
    Mining processes of Bitcoin and similar cryptocurrencies are currently incentivized with voluntary transaction fees and fixed block rewards which will halve gradually to zero. In the setting where optional and arbitrary transaction fee becomes the prominent/remaining incentive, Carlsten et al.\ [CCS~2016] find that an undercutting attack can become the equilibrium strategy for miners. In undercutting, the attacker deliberately forks an existing chain by leaving wealthy transactions unclaimed to attract petty complaint miners to its fork. We observe that two simplifying assumptions in [CCS~2016] of fees arriving at fixed rates and miners collecting {\em all} accumulated fees regardless of block size limit are often infeasible in practice and find that they are inaccurately inflating the profitability of undercutting. Studying Bitcoin and Monero blockchain data, we find that the fees deliberately left out by an undercutter may not be attractive to other miners (hence to the attacker itself): the deliberately left out transactions may not fit into a new block without ``squeezing out'' some other to-be transactions, and thus claimable fees in the next round cannot be raised arbitrarily. 

    This work views undercutting and shifting among chains rationally as mining strategies of rational miners. We model profitability of undercutting strategy with block size limit present, which bounds the claimable fees in a round and gives rise to a pending (cushion) transaction set. In the proposed model, we first identify the conditions necessary to make undercutting profitable. We then present an easy-to-deploy defense against undercutting by selectively assembling transactions into the new block to invalidate the identified conditions. Indeed, under a typical setting with undercutters present, applying this avoidance technique is a Nash Equilibrium. Finally, we complement the above analytical results with an experimental analysis using both artificial data of normally distributed fee rates and actual transactions in Bitcoin and Monero. 
\end{abstract}

\section{Introduction}\label{sec:intro}

Bitcoin network~\cite{nakamoto2008bitcoin} and several cryptocurrencies rely on nodes participating in transaction verification, ordering and execution, and mining new blocks for their security and performance. Specifically, with honest majority, Byzantine-fault tolerant consensus is possible with Proof of Work (PoW) assuming network synchrony. With honest majority, attacks like double spending~\cite{rosenfeld2014analysis} are also harder to implement in practice. Additionally, with more honest computing peers, liveness is provided with a higher probability. A proper incentive design helps attract more honest parties to join. 
Bitcoin currently incentivizes nodes (or miners) with fixed block rewards and voluntary transaction fees. Historically, the block reward has been the dominating source of miners' revenues. However, for Bitcoin, it is a system parameter that halves approximately every four years.\footnote{The next halving event to $3.125$ BTC is scheduled for May 2024. \cite{btchalving2}}
Its domination is expected to vanish due to the deteriorating nature and transaction fees will then become the major mining revenue generator. 

With a stable reward, a miner's expected revenues rely mostly on its probability of finding a block, which itself is contingent on the miner's hash power. However, in the fee-based incentive system, the revenues additionally depend on the fee amount inside a block, which further relies on users' offerings and miners' transaction selections. The total fees inside blocks are market-dependent and time-variant because (i) transaction arrival can be arbitrary; (ii) transaction fees are voluntary under the current mechanism, so they can be arbitrary (even 0) and the threshold fee rates for faster confirmation change with supply and demand in the block space market; (iii) miners have the freedom of sampling transactions to form new blocks. As a result, the fair sharing of revenue based on hashing power may not be maintained. 
For example, consider two miners $A$ and $B$ in the system with the same mining power. If $A$ mines blocks each with total fees of $1$~BTC and $B$ always encounters wealthy transactions and mines blocks each with $2$~BTC total fees, $B$'s revenue is twice $A$'s revenue. 

In particular, the fee-based incentivization framework nurtures a possible new deviating mining strategy called undercutting~\cite{carlsten2016instability}. 
In undercutting, the attacker intentionally forks an existing chain by leaving wealthier transactions out in its new block to attract other (petty compliant) miners to join the fork. 
Unlike honest miners, who follow the longest chain that appears first, petty compliant (PC) miners break ties by selecting the chain that leaves out the most fees. In \cite{carlsten2016instability}, fees accumulate at a fixed rate and miners claim \emph{all} accumulated fees when creating a new block. Thus, a miner undercuts another miner's block because it receives 0 of the fees in the target block but expects nonzero returns via forking. Similarly, PC miners join the fork because the undercutter leaves out more fees unclaimed (and they can claim \emph{all} fees in the next block). 
Carlsten et al. find that undercutting can become the equilibrium strategy for miners, thus making the system unstable as miners undercut each other.

However, this result is based on a setting disregarding the block size limit. If the fees claimable in the next block are bounded and a pending transaction set exists due to the block size cap, PC miners may not join the fork and undercutting may not be more profitable than extending the current chain head. The intuition is that the extra claimable fees are bounded, and the fork does not win with absolute probability, while the main chain may provide slightly fewer fees but extends with probability 1 when there's no attack. 
We give an illustrative example below where undercutting is not rational when we consider the limit. 
Let there be $33\%$ honest, $17\%$ undercutter, and $50\%$ PC mining power, $100$ total token fees with $20$ claimable in each block. 
The undercutter expects $3.4$ token returns by extending the chain head. 
Suppose it instead undercuts and claims half of the tokens in the target block, $10$ tokens, in its first forking block (as in~\cite{carlsten2016instability}). If PC miners do not shift, they expect ($0.5/0.83*0.83*20=$) $10$ tokens from the next main-chain block; if they follow the fork, they expect to gain ($0.5/0.67*0.67*20=$) $10$ tokens. But, shifting is not rational for the owner of the undercutting target block and may not be rational for others as they have started mining the main chain for some time. Even if they shift, we find the undercutter's expected return to be ($0.17*(0.67*10+0.67*0.17/0.67*20)=$) $1.717<3.4$. 

Towards modeling undercutting attacks more realistically and generally, we construct a new model to capture rational behaviors related to and performance of the undercutting strategy. Miners in our model are either honest or rational. A rational miner may \textbf{undercut} or \textbf{arbitrarily shift among chains} as long as the action maximizes its returns. Fees in our model arrive with transactions. By sorting transactions in the unconfirmed transaction set and packing at most a block size limit of transactions, we obtain the maximum claimable fees at a certain timestamp. Miners can choose to claim no more than this maximum fee.

Essentially, when undercutting, the rational miner's goal is to earn more than what it can potentially gain not undercutting. The attacker needs to first \textbf{(i)} attract other rational miners to join its fork if necessary, and second \textbf{(ii)} avoid being undercut by others. 
If it leaves out too many fees, it may end up being worse off undercutting. If it claims more than necessary, other rational miners may undercut its fork, annihilating its efforts. Then how many fees should an undercutter take to achieve both goals simultaneously? And can others make it not possible to do so? 
We seek to first locate such a feasible area for an undercutter to secure its premiums and next, uncover defenses against this attack. Note that undercutting is not desired because it hurts the expected profits for honest miners. Successful undercutting also harms users who attach high fee rates to have their transactions processed faster.

\subsection{Contributions}
We define an analytical model that captures behaviors that are ``rational'' but not necessarily ``honest'' like undercutting and shifting rationally. This can be used to analyze other rational deviating strategies in a fee-based incentive system. The key is to pinpoint reward distributions and probabilities of earning the rewards. 

Specifically for undercutting and as a key contribution, we offer \textbf{closed-form conditions on the unconfirmed transaction set to make undercutting profitable}. The key quantity is the ratio ($\gamma$) between the maximum claimable fees in the next block (w.r.t. block size limit) and the fees in the current block. For clarity, let the mining power fraction of the undercutter be $\beta_u$ and that of the honest miner be $\beta_h$, remaining rational miner be $\beta_r$. \textbf{(i)} In the best case for the undercutter in our model, the undercutter forgoes the fork after being one block behind instead of hanging on longer. \textbf{(ii)} When $\gamma < \frac{a \beta_r +\beta_u}{1-\beta_u}$, the attacker can expect to earn a premium by proper undercutting. It should carefully craft the first block on its fork (deciding parameter $a$) in such a way that rational miners can be attracted to join the fork when needed but not tempted to undercut it again. We provide more details in Section~\ref{theo2}. The conditions for the case where the undercutter holds on for one more block (\Cref{sec:sd2}) are stricter, as noted in (i) and the overall returns are worse. 

As a side-product and naturally, we provide an \textbf{alternative transaction selection rule} to counter undercutting, other than fitting all available transactions into a block. Once we have identified effective conditions for profitable undercutting, we work backward to proactively check the conditions before creating a new block. By making the conditions no longer satisfied, potential undercutters are no longer motivated to undercut. Applying the defense technique is Nash equilibrium in a typical setting. 
In the equilibrium, we additionally calculate the price of anarchy (PoA) to capture the inefficiency a strong undercutter brings or the advantage it has in a system. 
To make the system more stable, we can either strengthen the second potential undercutter or weaken the strongest undercutter through decentralization. 

We \textbf{experiment with real-world data} from Bitcoin and Monero blockchains to evaluate the profitability of undercutting and the effectiveness of avoidance techniques. We decide on the two systems because Bitcoin is representative of swamped blockchains and Monero typically has a small unconfirmed transaction set. 
\textbf{(i)} In Bitcoin, for a 17.6\% undercutter, the average return is 17.9\%. For a hypothetical 49.9\% attacker, the average revenue is 60.8\%. 
In Monero, we observe a profit increase of around 8 percentage points from fair shares for a 35\% attacker. 
\textbf{(ii)} After enabling defense, undercutting generates around a fair share for Monero 35\% undercutter where the two strongest rational miners possess the same mining powers. We test a strong undercutter's advantage in Bitcoin (49.9\%, 20\%), which gives the 49.9\% attacker around 63.5\% of the total returns. 

\paragraph{Organization of the paper} The rest of the paper is organized as follows: In \Cref{sec:Related}, we visit undercutting literature and related economic theories. \cref{sec:Def} offers a brief overview of blockchain mempools and mining and defines relevant concepts. In \Cref{theo} and \ref{theo2}, we model the mining game with undercutting and include an analytical study, while in \cref{sec:Exp}, we evaluate the profitability of undercutting and the effectiveness of the avoidance technique using real-world blockchain data. Finally, \Cref{sec:Conclusion} concludes the discussion.
\section{Related Work}\label{sec:Related}
Carlsten et al.\ \cite{carlsten2016instability} introduce the undercutting mining strategy to show the instability of the future Bitcoin fee-based incentivization system because undercutting can become the equilibrium strategy. There, transaction fees accumulate at a constant rate and miners can include all fees when creating a new block. But fees essentially are \textit{not} independent of transactions. If we dive into the transaction level and account for the block size limit, the fees one can claim are restricted and there can potentially be a large pending transaction set, which can cushion or even annihilate the effects of undercutting. 
Based on this intuition, we construct the new model focusing on transaction selection rules, which determine fees claimed and left out. Further, both undercutting and hopping among chains are modeled more generally as actions of rational miners instead of separately as two types of miners as in~\cite{carlsten2016instability}. This helps quantify the profit margin and brings about opportunities for mitigation. 

\paragraph{Together with other non-compliant mining strategies} There have already been rigorous discussions on attacks related to mining strategies. Most notable attacks are selfish mining~\cite{eyal2014majority,sapirshtein2016optimal,nayak2016stubborn}, block withholding~\cite{rosenfeld2011analysis,luu2015demystifying,courtois2014subversive,luu2015power,eyal2015miner}, and fork after withholding~\cite{kwon2017selfish}. Defenses against these game-theoretic attacks have also been studied~\cite{heilman2014one,zhang2017publish,pass2017fruitchains,kwon2019eye,lavi2019redesigning}. 
It is possible to combine undercutting with other mining strategies like selfish mining and block withholding. For the latter, because undercutters prefer larger mining power, the two attacks have opposite goals, so one needs to balance the computation resource allocation. Selfish mining purposely hides discovered blocks, while undercutting intends to publish a block and attract other miners. They do not share the same rationale, but we can schedule the two strategies and apply the one with higher expected returns at a certain time. In this work, we put our focus on the profitability and mitigation of undercutting, which affects the undercutting part of the strategy scheduler.

\paragraph{Sunk Cost} In traditional microeconomics \cite{mankiw2014principles}, a rational agent makes decisions based on prospective costs and disregard sunk costs. In behavioral economics \cite{benartzi1995myopic,tom2007neural}, decision-makers can have an irrational bias toward the probability distribution of future events, loss aversion, and other illusions. When a miner decides whether to continue on a chain or shift to other chains, it can be influenced by sunk costs including time already spent. We capture this mindset by letting rational miners shift after their current chain is $D\geq 1$ block(s) behind. A larger $D$ indicates a greater influence from sunk costs. 

\paragraph{Lemon Market} Another angle to look at the problem on a higher level is through the market for ``lemons''~\cite{akerlof1978market}, the brand-new car that becomes defective the minute one bought it. In the Bitcoin block space market, users are bidders, and miners are sellers. Users decide prices to pay based on their observation of the relationship between confirmation time and fee rates. They attach fee rates corresponding to the desired waiting time. If undercutting is prevailing, users who attach high fee rates but are ghosted are provided with ``lemons'' instead of ``peaches'' -- fast confirmation. This can result in a decrease in the overall fee rates, diminishing the profitability of undercutting. 
\section{Preliminaries}
\label{sec:Def}
\paragraph{Mempool}
Mempool~\cite{mempool} is an unconfirmed transaction set maintained by miners locally. When a transaction is announced to the network, it enters into miners' mempools. 
Miners select transactions from their mempools to form new blocks. Usually, a miner chooses the bandwidth set (Definition~\ref{bandwidthset}) with respect to the local mempool and global block size limit. 
An undercutting miner intentionally leaves out wealthy transactions when forming blocks to attract other rational miners. Wealthy transactions are those with high fee rates. 
When a new block is published, miners verify the block and then update their local mempools to exclude transactions included in the newly published block.

\begin{definition}[Bandwidth Set]\label{bandwidthset}
Given block size limit $B$ and an unconfirmed transaction set \mempool \ comprising $N$ transactions, $S^*\in P(\mathbb{T})$ is a bandwidth set of \mempool \ with respect to $B$ if $S^*.size \leq B$ and $\forall S_i\in P(\mathbb{T})$ with $S_i.size\leq B, S^*.fee\geq S_i.fee$, where $P(\mathbb{T})$ is the power set of \mempool. 
\end{definition}
\begin{remark}
A bandwidth set is a set of transactions in a miner's mempool providing the most fees a miner claimable in one block. 
If the unconfirmed transaction set is of size $\leq B$, then the bandwidth set is the memory pool itself. Note that the bandwidth set is not necessarily unique. 
\end{remark}

\begin{definition}[Safe margin]
When a chain $C^*$ is $D$ block(s) ahead of competing chains, a miner with safe margin parameter $D$ always extends $C^*$.
\end{definition}\label{safemargin}
\begin{remark}
Honest miners apply the longest chain rule and always have $D=1$.\footnote{When there is a tie, they choose the chain with the oldest timestamp. If timestamps should be the same, they select a chain at random.} For rational miners, $D\geq 1$. When the length discrepancy between competing chains is within $D$, they select the chain with the most expected returns. 
\end{remark}

\section{Mining Game Featuring Undercutting Strategy}\label{theo}
In this section, we model the mining game in the presence of undercutting attacks. We proceed with (i) mempools with "sufficient" unconfirmed transactions and (ii) mempools with "limited" transactions. We also allow the undercutter to apply two different safe depths $D=1$ and $D=2$, where it gives up attacking after being one or two blocks behind the main chain. For higher $D$, conditions for profitable undercutting become tighter, and overall its performance worsens.

In this section, we model the mining game involving the undercutting strategy. We consider honest miners, who follow the default protocol specifications, and rational miners. The latter are addressed as undercutters when they undercut. 

\paragraph{Game definition}\label{gamedesp}
We define the mining game $G=\langle M, A, R\rangle$ as follows:
\begin{itemize}[leftmargin=*]
 \item $n$ Players $M=\{M_0, M_1, ..., M_{n-1}\}$: without loss of generality, we label a subset of the miners that have a total of $\beta_h$ mining power as honest; we label a miner with $\beta_u$ mining power as the current undercutter under discussion; we label the remaining miners as (currently) non-undercutting rational miners and their total mining power is denoted as $\beta_r=1-\beta_h-\beta_u$. Honest miners are treated as one because they follow the same mining rules, and we assume they are informed the same way. 
 \item Actions $A=\{undercut(\cdot), stay(\cdot), shift(\cdot)\}$: we index chains during a game according to their timestamps after the branching point, e.g. the original (main) chain with index $Chain_0$, abbreviated as $C_0$. Honest miners always honest mine and may choose to stay or shift depending on circumstances. Rational miners may choose to undercut an existing chain and start a new chain, stay on a working chain, or shift among existing chains. 
 \item Utility functions $U=\{u_i\}_{M_i\in M}$: we let $u_i=R_i-c_i$, where $R_i$ is the total transaction fees it receives and $c_i$ is the cost. We treat the cost $c_i$ as fixed and reduce the problem of maximizing utility to maximization of obtained fees.
\end{itemize}

\paragraph{Threat model}
We allow no miner to own more than 50\% mining power (i.e., $\beta_u\leq 0.5$). We let miners publish their discovered blocks immediately to attract other miners to join. 
We assume the best case for the undercutter and let the mempool be the same for miners on the same chain. Because undercutting is not practical or meaningful if miners have distinct mempools, since wealthy transactions an attacker left unclaimed may not exist in others' mempools in the first place. This assumption makes the attacker stronger, and we intend to uncover what the attacker can obtain in advantageous environment settings. 

We let miners know of other miners' types (e.g. honest or rational) after sufficient observations. We assume miners can approximate the amount of mining power concentrated on a chain based on the block generation time on that chain. 

\paragraph{Solution concept} We solve for Nash Equilibrium (NE) in the mining game with the undercutting mining strategy. In a Nash Equilibrium, players do not earn extra utility by unilaterally deviating from the equilibrium strategy. 

\subsection{Miner's Winning Probability}\label{winprob}
A miner's expected returns from mining equal the product of its winning probability of a block and the fees residing in that block. Firstly, miner $M_i$'s winning probability of a block is simply its mining power when there is only one chain. In the case of competing chains, we need to additionally quantify a chain's winning probability when working in systems where only one chain survives.

\paragraph{A chain's winning probability} In undercutting, the attacker forks an existing chain by leaving out wealthy transactions. 
In the following discussions, we refer to the undercutting chain as $C_1$ and the current main chain as $C_0$. $C_0$ might not be on the main chain eventually if $C_1$ wins the race. The effective height of a chain is the number of blocks it has accumulated after the forking point. These competing blocks are called effective blocks in the game analysis.

Overall, the process proceeds as follows. The undercutter sees a new block is appended to $C_0$ by another miner. It starts to work on a forking block that excludes wealthy transactions appearing in the current chain head. With some probability, it can create the fork faster than the next block appearing on $C_0$. When the undercutter publishes its block, some rational miners consider shifting to $C_1$ because there are more high fee rate transactions that they can benefit from. To model this procedure, we screenshot the state of the system as a tuple that we denote as $\vec{S}=(m_0, m_1, \vec{F^0}, \vec{F^1}, O, \delta, \lambda_0, \lambda_1)$, where $m_0$ and $m_1$ are respectively the effective height of $C_0$ and $C_1$; $\vec{F^0}$ and $\vec{F^1}$ are the list of transaction fee total in effective blocks on $C_0$ and $C_1$; $O$ is the mining power currently working on $C_1$, which updates upon new block appending events; $\delta\in (-1,1)$ is the mining power shifting from the source chain to the destination chain, which is defined to be positive if miners are shifting to $C_1$ and negative if they are shifting to $C_0$; $\lambda_0$ and $\lambda_1$ are block generation rates for $C_0$ and $C_1$.

To obtain the winning probability measure for a chain from state $\vec{S}$, we view the block generation event as a Poisson process and use a random variable to represent the waiting time between block occurrence events. We denote waiting time for $C_0$ as $X$ and $C_1$ as $Y$. They both follow exponential distribution but with different rates. The rate parameters depend on the mining power distribution. 
Given the state $\vec{S}$, we obtain the block occurrence rate as: $ \lambda_0=\frac{1-O}{I};$ and $\lambda_1=\frac{O}{I}$, 
where $I$ is block generation interval (e.g. 10 minutes for Bitcoin). 
This is derived from the thinning theorem of the Poisson point process. The main idea is that independent sub-processes of a Poisson process are still Poisson processes with individual rates. With this property, we can determine the time interval for the next block to appear on a chain. Then, the key is the mining power concentrated on a chain, and further is whether honest and rational miners shift. 

For $D=1$, there is only one state that the currently non-undercutting rational miners $\beta_r$ need to make a decision, when the undercutter extends $C_1$ before the $C_0$ extends by one. The two competing chains are in a tie with relative height difference $\tilde{D}=0$. The probability that $C_1$ wins is simply $ p = \Pr[C_1\; Wins]= \Pr[Y<X]=O+\delta$.

For $D=2$, there is an infinite number of states where flexible rational miners need to make decisions about shifting. 
We let $\tilde{D}=m_1 - m_0<D$, denoting the number of blocks by which $C_1$ leads $C_0$. For example, when $\tilde{D}=-1$, $C_1$ is one block behind $C_0$. Then $C_1$ wins if it creates 3 blocks before $C_0$ extends by 1, or discovers 4 blocks before $C_0$ extends by 2, and so on. Thus, we have $p =\sum_{i=0}^{\infty}\Pr[(D-\tilde{D}+i) Y<(i+1) X]$.

\textbf{(i)}\textbf{When} $\mathbf{\tilde{D}=-1}$, $C_1$ is behind $C_0$. For $C_1$ to win, we need $p=\sum_{i=0}^{\infty}\Pr[(3+i)Y<(1+i)X]=\sum_{i=0}^{\infty}(\beta_u+\delta)^{3+i}(1-\beta_u-\delta)^{i}$. 

\textbf{(ii)}\textbf{When} $\mathbf{\tilde{D}=0}$, there is a tie between $C_1$ and $C_0$. In this case, $p=\sum_{i=0}^{\infty}\Pr[(2+i)Y<(1+i)X]=\sum_{i=0}^{\infty}(\beta_u+\delta)^{2+i}(1-\beta_u-\delta)^{i}$. 

\textbf{(iii)}\textbf{When} $\mathbf{\tilde{D}=1}$, $C_1$ is leading. We have $p=\sum_{i=0}^{\infty}\Pr[(1+i)Y<(1+i)X]=\sum_{i=0}^{\infty}(\beta_u+\delta)^{1+i}(1-\beta_u-\delta)^{i}$.

\paragraph{A miner's probability of winning a block} Suppose a miner $M_i$ with $\beta_{M_i}$ mining power is mining on a chain $C_j$ with $\beta_{C_j}$ accumulated total mining power which has winning probability $p_{C_j}$. Then $M_i$'s winning probability is $\frac{\beta_{M_i}}{\beta_{C_j}}p_{C_j}$.
\section{Game Analysis}\label{theo2}
\subsection{Giving Up After One Block Behind}\label{sec:sd1}
Now we discuss $D=1$. We use the abbreviated state $S^*=(m_0, m_1)$ in discussion. We denote the transaction fees inside the first two blocks of $C_0$ as $F^0_1$ and $F^0_2$, the transaction fees inside blocks of $C_1$ as $F^1_1$ and $F^1_2$, the expected returns for flexible rational miners $\beta_r$ as $R_r$ and the expected returns for the undercutter as $R_u$. When there is no undercutting, we denote their respective expected return as $R'_r$ and $R_{\overline{u}}$.

\begin{figure}
 \centering
 \includegraphics[scale=0.45]{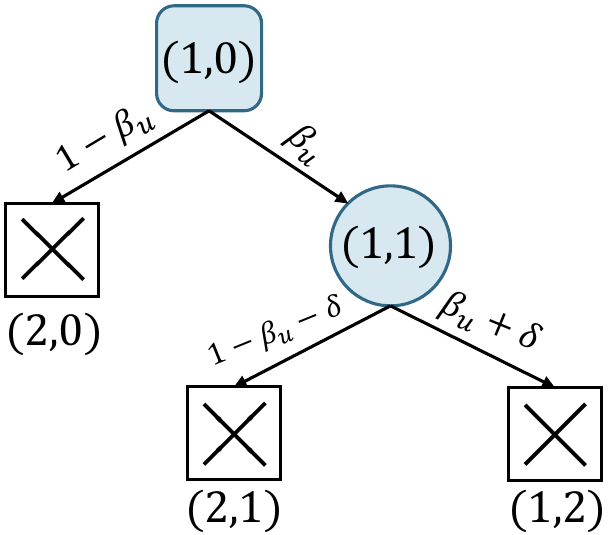}
 \caption{State transition for $D=1$. ``X'' Boxes are terminal states. For non-terminal states, circles indicate ties. Every left branch means $C_0$ extends by one and every right branch refers to $C_1$ creating a new block. The quantity on the arrow is the probability of state transition. Here $\delta$ is the amount of rational mining power shifting to $C_1$.}
 \label{fig:safemargin1}
\end{figure}

For $D=1$, rational miners only need to decide whether to shift at state $S^*=(1,1)$ when undercutting becomes visible as shown in Figure~\ref{fig:safemargin1}. 
Suppose they shift $x$ of their mining power $\beta_r$ to $C_1$. They can solve for $x$ in
\begin{align*}\label{shift3-ext}
 &\argmax E[R_r] = \argmax_{x\in [0,1]}\big( \mathbbm{1}_{owner}\cdot  (1-p)\cdot F^0_1 \\
 &+ \frac{(1-x)\beta_r}{\beta_h+(1-x)\beta_r} (1-p) \cdot F^0_2 +
 \frac{x\beta_r}{x\beta_r+\beta_u} p\cdot F^1_2\big)\numberthis{}
\end{align*}
where $p$ is the probability of $C_1$ winning and $\mathbbm{1}_{owner}$ indicates whether a rational miner is the owner of the first block on chain 0. The shift can then be calculated as $\delta = x\beta_r$. Observe that the optimization problem involves fees inside succeeding blocks after the forking point. We represent fees in a relative way for general interpretability: we let $F^0_1=1$ and have fee total in other blocks measured relative to it. 
Now we discuss two different mempool conditions.

\subsubsection{Mempools with limited bandwidth set} By ``limited'' we mean the current bandwidth set on $C_0$ has a small enough transaction fee total ($< \frac{\beta_u}{1-\beta_u} F^0_1$). We provide more details concerning this threshold as we proceed. WLOG, we assume $F^0_1=1$, $F^0_2=\gamma\geq 0$ (s.t. $\frac{F^0_2}{F^0_1}=\gamma$), $F^1_1=a$ and $F^1_2=b$ where $a\in [0,1]$. We can let $b=1+\gamma-a$, assuming the best case for the undercutter that it can compose the first block on $C_1$ in such a way that the second block can claim all unclaimed fees within one block. If a rational miner decides to undercut, with probability $\beta_u$, the undercutter can create a new chain and the game is started. 
In the remaining game, with probability $p=\beta_u+\delta$, $C_1$ wins and with probability $(1-p)$, $C_0$ wins. The expected profit of the undercutter is 
\begin{align*}
 E[R_u]=\beta_u(\beta_u+\delta)\cdot (1\cdot a + \frac{\beta_u}{\beta_u+\delta}\cdot (1+\gamma -a))
\end{align*}
The expected return for the rational miner if it does not undercut is $E[R_{\overline{u}}]=\beta_u\gamma$. 

The miner will undercut only if $E[R_{\overline{u}}]<E[R_u]$. Then 
\begin{equation}
 \gamma < \frac{\delta a+\beta_u}{1-\beta_u} \label{eq:gammacondition}
\end{equation}
With $\gamma < \frac{\beta_u}{1-\beta_u}$, $E[R_{\overline{u}}]<E[R_u]$ even when $\delta=0$. That is, even no rational miner shifts to $C_1$, there are so few fees left in the mempool that the attacker is always better off by forking $C_0$ compared with extending it. 

One extreme case is when there are no transactions left or the bandwidth set has negligible fees and $F^0_2=0$. 
The rational miner will fork because originally there is nothing left on $C_0$ and $E[R_{\overline{u}}]=0$. One detail is that the attacker needs to craft the first block (determine $a$) it generates to avoid being undercut again. Suppose when $\gamma < T$ ($T= \frac{\beta_u}{1-\beta_u}$ in our current context), a potential undercutter initiates the attack. Then by choosing $a$ in such a way that $\frac{1+\gamma-a}{a}\geq T'$ ($T'= \frac{\beta_{u_2}}{1-\beta_{u_2}}$ in the current context), the undercutter can avoid being undercut again. Note that here when an undercutter decides $a$, it is picturing a potential undercutter $\beta_{u_2}$ other than itself. We will revisit the choice of $a$ after complete the discussion for $\gamma > \frac{\beta_u}{1-\beta_u}$ case.

In conclusion, for $D=1$, when the attacker is stronger ($\beta_u$ is larger), the requirements on the mempool bandwidth set fee total for undercutting to be profitable regardless of rational miners' actions is looser. When $\beta_u$ approximates 0.5, the threshold ratio approaches 1, which occurs with high frequency. For $\beta_u=0.2$, the upper bound is 0.25, where the current bandwidth set is 1/4 of the fees inside the chain head of $C_0$.

\subsubsection{Mempools with sufficient bandwidth set} By ``sufficient'' we mean the current bandwidth set in the mempool has more than ``limited'' transaction fee total ($\geq \frac{\beta_u}{1-\beta_u} F^0_1$). In this case, the undercutter needs to attract some rational miners at state (1,1) (make $\delta>0$). It's straightforward to verify that the owner of the undercutting target block is better off by staying on $C_0$. We treat this miner as honest in the following calculations and only make decisions for the remaining rational players. 
To decide whether to shift to $C_1$, rational miners solve for $x$ in 
\begin{align*}\label{shift1}
 \argmax_{x\in [0,1]} E[R_r]=&\argmax_{x\in [0,1]}\big( \frac{(1-x)\beta_r}{\beta_h+(1-x)\beta_r} (1-p) \gamma \\
 &+\frac{x\beta_r}{x\beta_r+\beta_u} p (1+\gamma -a) \big)\numberthis{}
\end{align*}
Here $p=O+\delta=\beta_u+x\beta_r$. One observation is that the rational miners either move to $C_1$ with all their mining power or none (function is linear in $x$ after simplification). When $x=1$, we have $E[R_{r|x=1}]=\beta_r(1+\gamma -a)$. 
Similarly, in setting $x=0$, we obtain $E[R_{r|x=0}] =\beta_r \gamma$. 
To encourage shifting of rational miners, we need $E[R_{r|x=1}]>E[R_{r|x=0}]$, which means $a<1$. To avoid being undercut, the undercutter additionally needs to pick an $a$ such that this condition is not satisfied for the first block on its $C_1$. This is to say the undercutter can profitably undercut $C_0$ in expectation, but others do not expect to attack its $C_1$ successfully. As previously touched on, we need 
\begin{equation}\label{eq:avoidundercut}
 a\leq \frac{1+\bar{\gamma}}{1+T} =\frac{1+\bar{\gamma}}{1+\frac{ a_2 \beta_{r_2} +\beta_{u_2} }{1-\beta_{u_2}}}, 
 a_2 \leq \frac{1+\bar{\gamma}'}{1+\frac{ a \beta_{r} +\beta_{u} }{1-\beta_{u}}}
\end{equation}
where $\beta_{u_2}$ is the mining power of the strongest potential undercutter for this attacker, $a_2$ is what this opponent would claim in the first block if he forks the undercutter's chain and $\beta_{r_2},\beta_{h_2}$ is the remaining flexible rational mining power and honest mining power in that case. Here, $\bar{\gamma}, \bar{\gamma}'$ are the fee totals in the respective next bandwidth set measured relative to the respective current bandwidth set, when the strongest and second strongest undercutters are making the attack decisions. 
We can easily solve for $a$ and $a'$ numerically given assignments for mining power distributions and the mempool (for computing $\bar{\gamma}, \bar{\gamma}'$ from bandwidth sets). A program for this task can be found here \cite{codesrc}. 

For $D=1$, in conclusion, if $\gamma < \frac{\beta_u}{1-\beta_u}$, the undercutter with mining power $\beta_u$ can expect a potentially profitable undercutting attack. Otherwise, the undercutter sets $a$, the fees to claim in the first block (measured relative to the fees in the target block), properly and undercuts if $\gamma < \frac{a \beta_r +\beta_u}{1-\beta_u}$ for a potentially profitable attack. We say ``potentially'' because new transactions may arrive and change the bandwidth set, resulting in uncertainties in implementing undercutting. We summarize below the algorithm for rational miners to decide whether to attack, how to distribute mining resources, and how to avoid being undercut.

\begin{tcolorbox}[breakable, enhanced]\label{sd1}
 \textbf{(Part 1)} A potential \textbf{undercutter} decides whether to undercut: 
 
 Compute $a$ numerically according to Inequalities~\ref{eq:avoidundercut} that maximizes $E[R_u]$ and check if $\gamma < \frac{a \beta_r +\beta_u }{1-\beta_u}$. If Yes, start undercutting.

 \textbf{(Part 2)} \textbf{Flexible rational miners} decide mining resource distribution: 

 Solve for $x$ (proportion of resources to shift to the chain) in Equation \ref{shift3-ext}.
 
 \textbf{(Part 3)} \textbf{Miners} avoid being undercut: 

 Calculate the attack condition $T$($=\frac{a \beta_r +\beta_u }{1-\beta_u}$) for the strongest undercutter a miner is defending against. Check if the current $\bar{\gamma}< T$. 
 If Yes, include in the current block $<\frac{1+\bar{\gamma}}{1+T} $ of the fees in the bandwidth set; otherwise, use the bandwidth set.
\end{tcolorbox}

\paragraph{Treating rational miners as a whole} In the above analysis, rational miners make decisions from a collective perspective by maximizing $E[R_r]$ instead of the expected returns for a specific rational miner. This can give rise to coordination problems. Fortunately, rational miners either move all their mining power or stay on their current chain. There is only one state $(1,1)$ where they need to make a decision. There is one scenario in practice when a rational miner may not be flexible, which is when this miner owns the current chain head of $C_0$. When a rational miner is not flexible, as mentioned in the above analysis, we treat it like an honest miner. 
Since miners are aware of other miners' types across time, they will be able to adjust their reasoning process. 

\paragraph{When to apply undercutting avoidance} Suppose the current bandwidth set contains fees of 1 and the remaining next bandwidth set contains fees of $\gamma$. The mempool is always sorted so $\gamma\leq 1$ (except when no transaction exists and $\gamma$ is not well-defined). Suppose we have computed the corresponding threshold attacking condition $T$ for a rational attacker and $\gamma <T$. Then this attacker undercuts if a miner simply assembles the current bandwidth set into a block or claims $\geq \frac{1+\gamma}{1+T}$ of the fees in the bandwidth set. We state the following theorem. 
\begin{restatable}[]{theorem}{done}
In setting $D=1$, each miner applying avoidance procedure when creating a new block is NE. 
\label{thm:d1}
\end{restatable}
\begin{proof}
Let $M_i\in M$ be a miner with mining power $\beta_{M_i}$ and $M_i$ calculates $T=\frac{ a\beta_r +\beta_u }{1-\beta_u}$. When $\gamma \geq T$, $M_i$ proceeds as normal. Therefore, we only need to show that for $M_i$, when $\gamma < T$, $M_i$ is better off by claiming $a<\frac{1+\gamma}{1+T} $ of the fees in bandwidth set. The key element here is that the decision of how many fees to claim in a block is decided before one successfully generates the proof of work. 
Let the current bandwidth set $BS_0$ have a fee total of 1, and we measure the expected returns relative to it. 
We denote $M_i$'s expected return from not applying avoidance as $E[R_{M_i}]$ and applying avoidance as $E[R_{M_i,avoid}]$. 

It's straightforward to see that $E[R_{M_i,avoid}]=1 \cdot \beta_{M_i} = \beta_{M_i}$ because the strongest and other rational miners do not undercut. $M_i$ can claim fees in the current bandwidth set $BS_0$ in different rounds. 
Each time, $M_i$ generates a successful proof of work with probability $\beta_{M_i}$. 

If $M_i$ does not apply avoidance and claim all fees in $BS_0$, at least the strongest rational miner is incentivized to undercut given that $\gamma < T$. From previous analysis (see \Cref{fig:safemargin1} for a quick reference), we know that the undercutter wins with probability $\beta_u(\beta_u+\delta)$ where $0\leq \delta \leq \beta_r - \beta_u$. Thus, $M_i$ can expect to gain profits $E[R_{M_i}]= 1 \cdot \beta_{M_i} (1-\beta_u(\beta_u +\delta)) < E[R_{M_i,avoid}]$. 

By unilaterally deviating from avoidance when $\gamma$ satisfies undercutting conditions of a potential undercutter, $M_i$ receives smaller expected returns. 
\end{proof}
There are two special cases worth noting: (1) all miners are honest ($\beta_h=1$) so that $T=0$. 
We know that $\gamma\geq 0$. No effective avoidance is ever needed in this case; (2) $M_i$ is the only rational miner ($\beta_r=0$) so that $T=0$ for itself. $M_i$ does not need to apply avoidance since $\gamma\geq 0$. 

\paragraph{Quantifying Strong Undercutter's Advantage} Let the strongest undercutter have mining power $\beta_u$ and the second strongest undercutter have mining power $\beta_{u_2}$. We know from the previous discussion that a miner should always apply avoidance techniques to avoid being undercut in our current setting. For miners other than the strongest undercutter $\beta_u$, they need to defend against $\beta_u$ while $\beta_u$ itself only needs to defend against $\beta_{u_2}$. Let $T,T'$ be the threshold ratio computed for $\beta_u$ and $\beta_{u_2}$ respectively. We can capture its advantage with the ratio $\frac{1+T}{1+T'}$. For example, if $\beta_u=0.5,\beta_{u_2}=0.2,\beta_h=0$, $\frac{1+T}{1+T'}=4$, which means that the strongest undercutter can claim 4 times than what the other miners are collecting each time. When the discrepancy between $\beta_u,\beta_{u_2}$ approaches 0, $\frac{1+T}{1+T'}$ approaches 1. More formally, we capture this inefficiency brought by selfish behavior with the price of anarchy (PoA)~\cite{koutsoupias1999worst}. 

\begin{corollary}[Price of Anarchy]
In setting $D=1, \beta_h<1, \beta_r>0$, with the strongest and the second-strongest undercutters respectively having mining power $\beta_u, \beta_{u_2}$, the Price of Anarchy is $PoA=\frac{1+T}{(T-T')\beta_u+1+T'}$, where $T,T'$ are as defined above. 
\end{corollary}
This follows from the above analysis. When all miners stay honest, the ``undercutter'' is expected to earn a fair share $\beta_u$. When miners apply avoidance, the strongest undercutter claims $\frac{1+\gamma}{1+T'}$ each time while others claim $\frac{1+\gamma}{1+T}$. We can obtain its share $\frac{\beta_u \frac{1+\gamma}{1+T'}}{\beta_u \frac{1+\gamma}{1+T'} + (1-\beta_u) \frac{1+\gamma}{1+T}}$. Then we can calculate the PoA as the ratio between the strongest undercutter's shares in its optimal situation (the worst-case NE for the system) and its worst case (the optimal all honest outcome). We do not include other miners' returns in the calculation because the total shares always sum up to 1 regardless of the outcome and our focus is on capturing the advantage of the undercutter. 
To give a demonstrative example, let $\beta_u=0.499, \beta_{u_2}=0.176$ and $\beta_h \in \{0, 0.05, 0.10,\ldots, 0.30 \}$, on average (over $\beta_h$) $T=1.30,T'=0.29$ and $PoA=1.29$. 
This means that for $\beta_u$, the mean revenue proportion from undercutting is $0.499\times 1.29=0.63$. 

One observation is that when $\beta_u$ and $T- T'$ are large, PoA is large. To move it towards 1 (a more stable system), we can either strengthen the second potential undercutter or downsize $\beta_u$ through further decentralization.

\subsection{Giving Up After Two Blocks Behind}\label{sec:sd2}
Now, we discuss $D=2$. Rational miners now make decisions at states $S^*=\{(1,1),$ $(1,2),(2,1),(2,2),...\}$. The probability $p$ now comprises infinite series. Without loss of generality, we let $F_{1}^0=1$, $F_2^0=F_3^0=\gamma$, $F_1^1=a,F_2^1=b$ and $F_3^1=1+2\gamma - a - b $ (where $a\in [0,1],\gamma\geq 0$). $F_2^0,F_3^0$ can be of different values in reality but here we use the same value to highlight the wealthiness of $F_1^0$. Suppose eventually we derive an attacking condition $T$ for setting $D=2$ as well, then the undercutter would want to set $a$ and $b$ to satisfy $\frac{1+\gamma-a}{a} > T$ and $\frac{1+2\gamma-a-b}{b} > T$ to avoid being undercut. 

\begin{figure}
 \centering
 \includegraphics[scale=0.45]{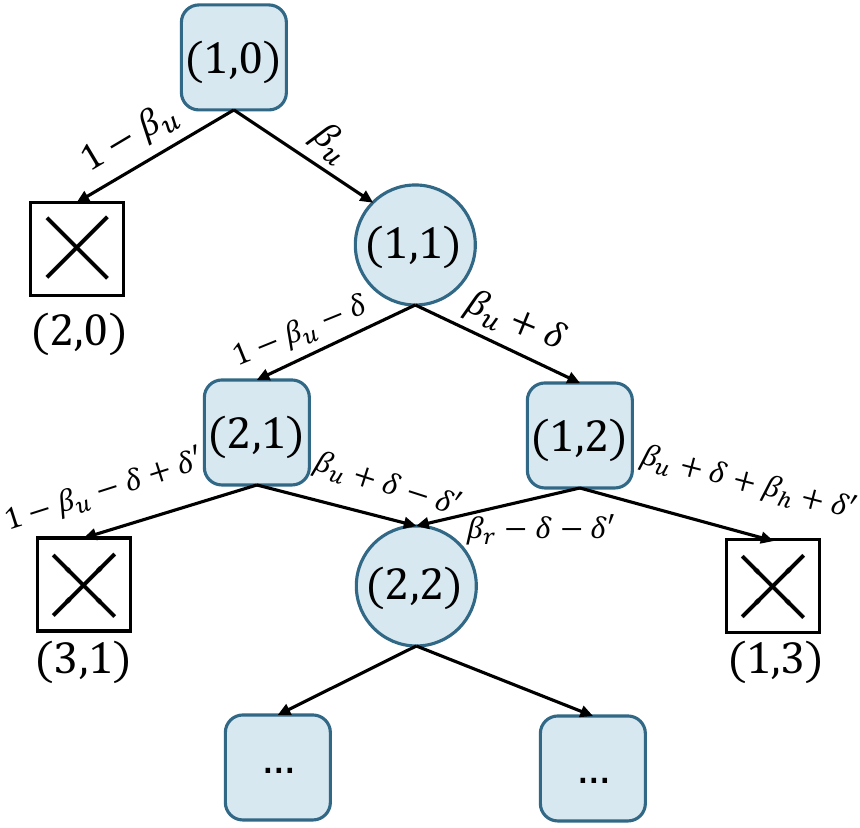}
 \caption{State transition for $D=2$. Notations are the same as Figure \ref{fig:safemargin1}. Now we have infinite state transitions. $\delta'$ and $\delta''$ are the amount of rational mining power shifting from one chain to another. 
 } 
 \label{fig:safemargin2}
\end{figure}

We take the same route as in the $D=1$ case. 
We know that if there is no attack, the undercutter expects to receive $E[R_{\overline{u}}]=2\beta_u\gamma$. 
If it starts the attack, its expected return from the right branches (shown in Figure~\ref{fig:safemargin2}) when the undercutter succeeds and no rational miners assist is 
\begin{align*}
 E[R_u]
 &=\beta_u(2\gamma+1)\sum_{i=0}^{\infty} \beta_u^{i+2}(1-\beta_u)^i =\beta_u^3(2\gamma+1) \\
 & \cdot \lim\limits_{n\rightarrow \infty} \frac{1-\beta_u^n (1-\beta_u)^n}{1-\beta_u(1-\beta_u)} = \frac{\beta_u^3(2\gamma+1)}{1-\beta_u(1-\beta_u)}
\end{align*}

When $\gamma <\frac{\beta_u^2}{2(1-\beta_u)}$ (with limited bandwidth set), the undercutter can expect to successfully start the attack without rational miners joining $C_1$. This bound is more demanding than the one for $D=1$. For $\beta_u=0.5$, the upper bound is now $0.25$ instead of 1. For $\beta_u=0.2$, the bound is 0.025 instead of 0.25. Overall, for weak attackers, the condition is way more demanding than before.

Next, we consider $\gamma \geq \frac{\beta_u^2}{2(1-\beta_u)}$ (with sufficient bandwidth set) and the undercutter needs rational miners to join $C_1$. Same as before, rational miners allocate their mining power among the two chains to maximize their expected returns:

\begin{align*}\label{shift2}
 \argmax_{x\in [0,1]} E[R_r] & = \argmax_{x\in [0,1]}\bigg( \mathbbm{1}_{owner} \cdot 
 p_0+ \frac{(1-x)\beta_r}{\beta_h+(1-x)\beta_r} p_0 \\ 
 & \cdot 2\gamma + \frac{x\beta_r}{x\beta_r+\beta_u} p_1\cdot b \\ 
 & +\frac{x\beta_r}{x\beta_r+\beta_u+\beta_h} p_1 \cdot (1+2\gamma -a-b) \bigg)\numberthis 
\end{align*}
where $p_0\leq (1-\beta_u-x\beta_r)^2$ is the probability of $C_0$ leading by 2 blocks first and $p_1\geq (\beta_u+x\beta_r)(\beta_u+x\beta_r+\beta_h)$ is the probability of $C_1$ leading by 2 blocks first. Here we only consider the leftmost and rightmost branches in Figure~\ref{fig:safemargin2} because they are the two most significant paths. We can observe that the objective function is convex. By Jensen's inequality, the expected returns reach maximum at either of the two ends. 
Again we let $E[R_{r|x=0}]<E[R_{r|x=1}]$ and obtain 

\begin{align*}
 &2(1-\beta_u)\gamma < b + (\beta_u+\beta_r) (1+2\gamma - a - b) \\
 &\stackrel{\beta_h > \beta_u}{\Rightarrow }
 \gamma < \frac{ (\beta_u + \beta_r) (1 - a ) + \beta_h b }{2(\beta_h - \beta_u)}
\end{align*}
When $\beta_h\leq \beta_u$, flexible rational miners move to the fork if $b>0$. 
With rational miners joining, the expected return for undercutter on the rightmost branch is now 
$$E[R_{u}] = \big ( a + \frac{\beta_u}{\beta_u+\beta_r} b + \beta_u (1+2\gamma - a - b )\big ) \cdot \beta_u(\beta_u+\beta_r)$$ 
We let $E[R_{u}] > E[R_{\overline{u}}]$ and obtain the condition on $\gamma$ for profitable undercutting while attracting other rational miners to join:
\begin{multline}
 \gamma < \min \{\frac{ (\beta_u + \beta_r) a + \beta_u b + \beta_u(\beta_u + \beta_r) (1-a-b) }{2 ( 1 - \beta_u (\beta_u + \beta_r) ) }, \\
 \mathbbm{1}^*_{\beta_h > \beta_u}
 \frac{ (\beta_u + \beta_r) (1 - a ) + \beta_h b }{2(\beta_h - \beta_u)} \} 
 \label{eq:dtwogamma}\numberthis
\end{multline}
where $\mathbbm{1}^*_{\beta_h > \beta_u} = \infty$ if $\beta_h \leq \beta_u$ and $1$ otherwise. 
Same as before, we denote the right-hand side condition as $T$ and solve for $a$ and $b$ numerically by considering the strongest potential undercutter the attacker is facing. 
\begin{equation}\label{eq:avoidundercutd2}
 a\leq \frac{1+\bar{\gamma}}{1+T}, 
 a_2 \leq \frac{1+\bar{\gamma}'}{1+ T'},
 b \leq \frac{1+2\tilde{\gamma}-a}{1+T}, 
 b_2 \leq \frac{1+2\tilde{\gamma}'-a}{1+T'}
\end{equation}
where $T$ and $T'$ are the attack conditions for the undercutter under discussion and its strongest opponent. 
Here, $\tilde{\gamma}, \tilde{\gamma}'$ are the fee totals in the respective third bandwidth set measured relative to the respective next bandwidth set.

In conclusion, for $D=2$, the limited bandwidth set bound is now $\gamma < \frac{\beta_u^2}{2(1-\beta_u)}$. This criterion can be hard to meet for weak miners with less than 30\% mining power ($\gamma <0.065$ for $\beta_u=0.3$). But for strong attackers with 40\%-50\% mining power (0.13 - 0.25), the conditions are not rare to satisfy. In the sufficient bandwidth set scenarios, attackers also have tighter bounds on $\gamma$ to initiate profitable attacks, especially for weak attackers. 
We present the algorithm for $D=2$ below.

\begin{tcolorbox}[breakable, enhanced]\label{sd2}
 \textbf{(Part 1)} A potential \textbf{undercutter} decides whether to undercut:

 Compute $a,b$ numerically according to Inequalities~\ref{eq:avoidundercutd2} that maximizes $E[R_u]$ and check if $\gamma$ satisfies Inequality~\ref{eq:dtwogamma}. If Yes, start undercutting.

 \textbf{(Part 2)} \textbf{Flexible rational miners} decide mining resource distribution:

 Solve for $x$ in Equation \ref{shift3}. 
 
 \textbf{(Part 3)} \textbf{Miners} avoid being undercut:

 Calculate the attack condition $T$ (right-hand side of Inequality~\ref{eq:dtwogamma}) for the strongest undercutter a miner is defending against. Check if current $\bar{\gamma}< T$. 
 If Yes, include in the current block $<\frac{1+\bar{\gamma}}{1+T} $ of the fees in the bandwidth set; otherwise, use the bandwidth set.
\end{tcolorbox}

We now give the general objective function for solving the mining resource allocation for flexible rational miners. 
Suppose $C_1$ extends by one, rational miners redistribute their mining power allocated on $C_0$ to $C_1$ by solving
\begin{align*}\label{shift3}
 & \argmax_{x\in [0,1]} E[R_r]=\argmax_{x\in [0,1]}\bigg( \text{(fees own on $C_0$)}p_0+\\ 
 & \text{(fees own on $C_1$)}p_1+\text{(claimable fees on $C_0$)}\frac{(1-x)\beta_r}{1-O-x\beta_r} p_0 \\
 & +\text{(claimable fees on $C_1$)}\frac{x\beta_r}{O+x\beta_r} p_1\bigg)\numberthis 
\end{align*}
where $p_1=(O+x\beta_{r}^0)^{D-\tilde{D}}$ and $p_0=(1-O-x\beta_{r}^0)^{D+\tilde{D}}$ ($\beta_{r}^0$ is the rational mining power on $C_0$). 
Claimable fees are total fees that a miner can expect to obtain in the unconfirmed transaction sets of each chain within size limit $(D\mp\tilde{D})\cdot B$. When $C_0$ extends by one, we have $p_1=(O-x\beta_{r}^1)^{D-\tilde{D}}$ and $p_0=(1-O-x\beta_{r}^1)^{D+\tilde{D}}$ ($\beta_{r}^1$ is the rational mining power on $C_1$).

\paragraph{Unobservable block owners}
Suppose one cannot observe the owner of blocks. Then we substitute $\mathbbm{1}_{owner}$ with $\frac{\beta_r}{\beta_r + \beta_h}$, the probability of a rational miner owning the target block conditioned on that the undercutter is not the owner. In $D=1$ setting, flexible rational players assist the undercutter if $a < \frac{\beta_h}{\beta_r + \beta_h}$. This is added to Equation \ref{eq:avoidundercut} as additional condition on $a,a_2$ when we solve for the two parameters numerically. For $D=2$, flexible rational players assist the undercutter if 
$$2(\beta_h - \beta_u)\gamma < \frac{\beta_h^2}{\beta_r+\beta_h} + \beta_u - \beta_h - (\beta_u + \beta_r) a + \beta_h b $$ 
This changes the second term on the right-hand side in Equation \ref{eq:dtwogamma} slightly.  

\subsection{Rearranging of Past Blocks in Extreme Case}\label{sec:drought}
Consider an extreme case where there are only negligible fees unclaimed in the unconfirmed transaction set for a sufficiently long time (greater than multiple block generation intervals $I$). As we have noticed, when there is only a limited amount of fees left in the mempool, a rational miner can avoid undercutting by claiming only a part of the bandwidth set. But if the situation continues to worsen and no new transactions enter the system, the remaining transaction fees become negligible at a certain point. In this extreme case, rational miners may look back to the previous blocks and start undercutting at certain block heights and rearrange blocks from there onwards. Suppose an undercutter goes back $C$ blocks. As long as there are only negligible transaction fees flowing into the system during $ \frac{C\cdot I}{\beta_u}$, it's more desirable for an attacker who earned less than its fair share in the past $C$ blocks to attack. The previous analysis does not apply to this extreme case. 
\section{System Evaluation}\label{sec:Exp}
In this section, we evaluate the profitability of undercutting using data obtained from Bitcoin and Monero, along with artificial transactions generated from normal distributions. Bitcoin is a typical example of congested blockchains, and Monero is a more available one. The simulation codes and a sample data set have been made open source~\cite{codesrc}. In the previous analysis, we let the undercutter be aware of future transaction flows in and out of the mempool. In reality, there is more uncertainty involved. 
Another difference is that now mining powers are discrete, and we model each miner individually.

\subsection{Data Collection}\label{sec:data_collcetion}

\paragraph{Transactions}
We obtained the blocks from height $630,457$ (May 15th, 2020 after the Bitcoin's block reward halving) 
to {$634,928$ (June 15th, 2020)} from the Bitcoin blockchain using the API provided by blockchain.com~\cite{blockchaindotcom}, comprising of $9,167,040$ transactions. 
Similarly, we further obtained data for the Monero blockchain using a similar API from \url{xmrchain.net}. In total, we obtained $1,482,296$ transactions from block height $2,100,000$ (May 17th, 2020) to $2,191,000$ (Sept 20th, 2020). 

For each of these transactions, we extracted the size, fee, and timestamp. 
The timestamp serves as a proxy for the time that the transaction arrives at the miners' mempool. Note that transactions that appeared during this time frame but not in any of the collected blocks are not included. Therefore, the memory pools reconstructed are not the exact mempools miners were faced with. We also create artificial transaction data sets with normally distributed fee rates.

\paragraph{Miners} There are three types of miners in the experiment. 
The \textit{undercutting} miner undercuts. 
The \textit{honest} miners follow the policies stated in the Bitcoin protocol, which is to extend the longest chain and break ties according to the block's broadcast time.
The \textit{rational} miners shift among chains to maximize their expected profits regardless of default rules. Conceptually the \textit{undercutting} miner is also a \textit{rational} miner. The largest rational miner is made to be the undercutter. 

To mimic the Bitcoin network's current state, we follow the mining power distribution of miners published by blockchain.com~\cite{miningpowers} on July 30th, 2020.
In total, we have $16$ miners, 
with mining powers ranging from $0.6$ to $17.6$ percent. To give the adversary the advantage, we select the strongest miner with $17.6\%$ of the mining power as the undercutting miner. The remaining 15 miners are distributed between honest and rational miners, as explained later. 
We additionally consider a hypothetical undercutter with 49.9\% mining power. This is to uncover the profitability of undercutting for a strong attacker and its advantage over other miners when avoidance techniques are adopted by all. 

For the Monero network, we follow the mining power distributions published by exodus~\cite{exodusmoneropools} and moneropool.com~\cite{moneropoolsdotcom}. The strongest pool with 35\% mining power is made the undercutting miner. 

\makeatletter
\newcommand{\removelatexerror}{\let\@latex@error\@gobble}
\makeatother

\begin{figure}[t]
\colorbox[gray]{0.95}{
\begin{minipage}{0.95\columnwidth}
\SetAlgoLined
\SetNlSty{textbf}{}{:}
\begingroup
\removelatexerror
\begin{algorithm}[H]
\small
 \SetKwInOut{Input}{input}\SetKwInOut{Output}{output}
 \SetAlgoLined
 \DontPrintSemicolon
 \Input{\texttt{txSet}, \texttt{minerSet}, \texttt{chainsTime}} 
 \While{\texttt{txSet} not empty}{
 extChain $\leftarrow$ nextChainToExtend(\texttt{chainsTime});
 
 m $\leftarrow$ selectNextBlockMiner(extChain);
 
 nextBlock $\leftarrow$ publishBlock(m);
 
 \vspace{1mm}
 updateChains(extChain, nextBlock);
 
 \vspace{1mm}
 updateMiners(extChain);
 
 \vspace{1mm}
 updateMempool(extChain);
 }
\caption{Simulation Overview}\label{alg:simulation} 
\end{algorithm}
\endgroup
\end{minipage}}

\colorbox[gray]{0.95}{
\begin{minipage}{0.95\columnwidth}
\SetAlgoLined
\SetNlSty{textbf}{}{:}
\begingroup
\removelatexerror
\begin{algorithm}[H]\small
 \DontPrintSemicolon
 \SetKwFunction{Fminers}{updateMiners}
 \SetKwFunction{Fchains}{updateChains}

 \SetKwProg{Fn}{Function}{:}{}
 \Fn{\Fchains{extChain, nextBlock}}{
 extChain.append(nextBlock);
 
 \ForEach{chain in \texttt{chainsTime}}
 {remove from \texttt{chainsTime} if it is non-wining}

 t $\leftarrow$ NextBlockCreationTime(extChain);
 
 update \texttt{chainsTime} with tuple (extChain, t);
 
 }
 \vspace{2mm}

 \SetKwProg{Pn}{Function}{:}{}
 \Pn{\Fminers{extChain}}{
 \ForEach{miner in minerSet}{
 
 \uIf{miner = undercutter}{decide to fork or not and craft the new block as described in Part 1 of the $D=1$ algorithm in \ref{sd1}, the $D=2$ algorithm in \ref{sd2};}
 
 \uIf{miner = honest}{\uIf{extChain longest chain}{switch to extChain;}}
 
 \uIf{miner = rational} {decide to switch to extChain or stay on current chain as described in Part 2 of the $D=1$ algorithm in \ref{sd1}, the $D=2$ algorithm in \ref{sd2};}
 }
 
 }
\caption{Chain and Miner Updates \label{alg:chain_miner_update} }
\end{algorithm}
\endgroup
\end{minipage}}

\end{figure}

\subsection{Experiment Setup}
We model the blockchain system as event-based, where the events are new block creations. Parameters and states of the system are updated upon creating a new block that we denote as $B_i$ for the remaining of this section. 
We assume that all miners have the same view of the network and the same latency in propagating the blocks and transactions. So miners working on the same chain have the same mempool.

\paragraph{Initial setup} 
We initialize the system's time to the earliest timestamp ($t_0$) of the collected transaction. Then we create the empty genesis block $B_0$ and create the chain $C_0$ by appending $B_0$ to it. Next, we insert the tuple ($C_0$, $t_0$) to an empty list \texttt{chainsTime}. The tuples inside this list indicate when the next block for each of the chains will be generated.
Alg. \ref{alg:simulation} provides an overview of the simulation after the initial setup. 
The simulation takes the transactions (\texttt{txSet}), miners (\texttt{minerSet}) and 
the tuple list (\texttt{chainsTime}) as inputs.
We consider these inputs as global for all functions in the simulation. Each iteration of the while loop indicates a new event.

\paragraph{Block creation (line 2-4 in Alg. \ref{alg:simulation})}
In the first step of each iteration, the chain to be extended, \texttt{extChain}, is selected using the \texttt{nextChainToExtend} function. It sorts all the tuples in \texttt{chainsTime} and picks the chain with the smallest next block creation time.
Next, the algorithm selects the new block's miner using the function \texttt{selectNextBlockMiner}. This function randomly selects miner \texttt{m}, from all the miners that are working on \texttt{extChain}, weighted by their mining power.
Finally, the selected miner \texttt{m} publishes the next block using the transactions in its mempool.

\paragraph{Chain updates (line 5 in Alg. \ref{alg:simulation})}
After the creation of the new block $B_i$, all the chains in the system are updated via the procedure depicted in Alg. \ref{alg:chain_miner_update}.
In the first step, block $B_i$ is appended to the current extending chain \texttt{extChain}.
Next, all other chains are checked against the \texttt{extChain} to see if they are in a non-winning situation (\texttt{extChain} has at least $D$ blocks from the forking point). 
If a chain is non-winning, it will be removed from the system, and \texttt{chainsTime} will be updated.
Finally, the \texttt{chainsTime} list is updated with the new time for the next block on \texttt{extChain}.

\paragraph{Miner updates (line 6 in Algo. \ref{alg:simulation})}
Following the chain updates, miners update their working chains.
Each miner based on its type (undercutter, honest, rational) decides whether to change its working chain (shown in \texttt{updateMiners} function in Algo.~\ref{alg:chain_miner_update}). 

\begin{enumerate}[leftmargin=*]
 \item If \texttt{extChain} is a competing chain of an undercutter miner, the miner checks whether \texttt{extChain} is $D$ blocks ahead of its own chain and switches to \texttt{extChain} if Yes. 
 
 \item If \texttt{extChain} is not a forked chain created by the undercutter, and miner \texttt{m} (miner of block $B_i$) is not the undercutting miner itself, the undercutter begins the condition checking routine. If the condition is ripe, it forks block $B_i$. Otherwise, it continues to extend \texttt{extChain}.

 \item All honest miners check whether \texttt{extChain} is the longest chain in the system. If Yes, they switch to chain \texttt{extChain}.
 
 \item Rational miners that are not on \texttt{extChain}, compare the length of \texttt{extChain} with their current chain and calculate their expected returns as described in \cref{sec:sd1,,sec:sd2}, and decide whether to switch to \texttt{extChain}.
\end{enumerate}

\paragraph{Mempool update (line 7 in Algo. \ref{alg:simulation})}
The last system update before moving to the creation of the next block ($B_{i+1}$) is the mempool update. 
In this step, all transactions in \texttt{txSet} with a timestamp between the creation time of $B_{i-1}$ and $B_i$ are added to the mempool of the miners on \texttt{extChain}.

\paragraph{Simulation run}
In a normal run, we repeat the above steps until all transactions have been consumed (\texttt{txSet} is empty). In an avoidance-enabled simulation run, we repeat the procedure but with all miners actively defending against undercutting in line 4, according to the two summarized algorithms in \Cref{sd1,,sd2}.
To moderate fluctuations caused by the random selections (of block generation time and block owners), we repeat the experiments 50 times (for each parameter set) for Bitcoin, 10 times for Monero, and report the mean values of profit proportions along with the 95\% confidence intervals.

\begin{figure*}[t]
 \centering
 \begin{subfigure}{0.45\textwidth}
 \centering
 \includegraphics[width=\columnwidth]{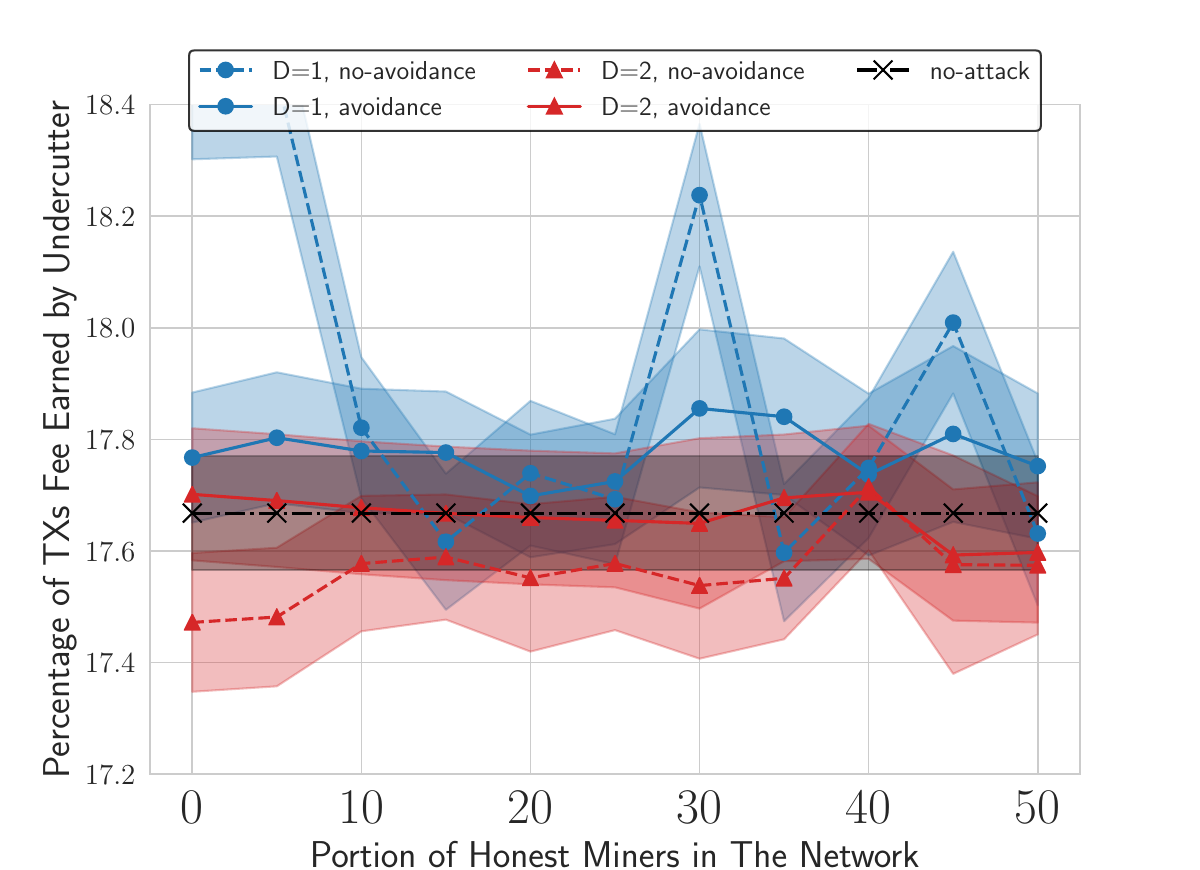}
 \caption{Bitcoin: Returns for $17.6\%$ Miner. }
 \label{fig:bitcoin17}
 \end{subfigure} 
 \begin{subfigure}{0.45\textwidth}
 \centering
 \includegraphics[width=\columnwidth]{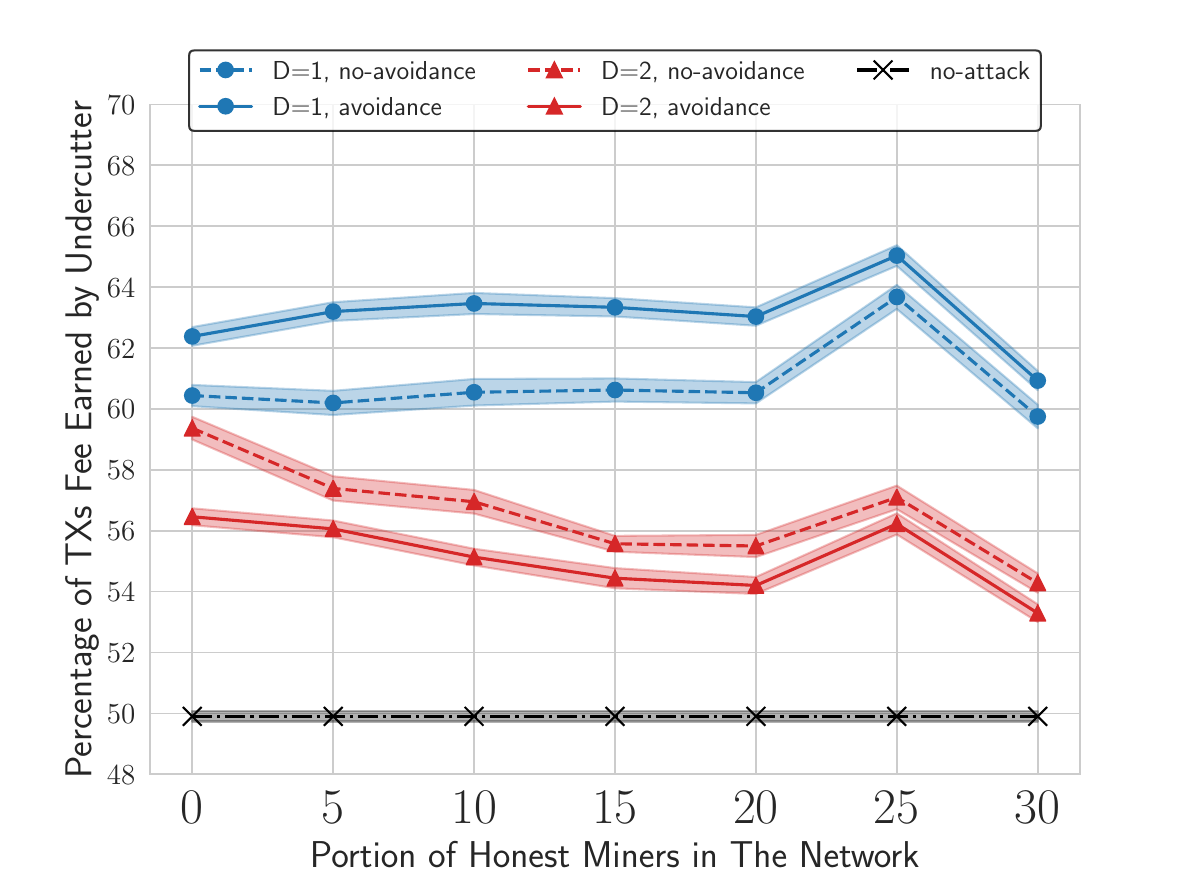}
 \caption{Bitcoin: Returns for $49.9\%$ Miner. }
 \label{fig:bitcoin49}
 \end{subfigure}
 \begin{subfigure}{0.45\textwidth}
 \centering
 \includegraphics[width=\columnwidth]{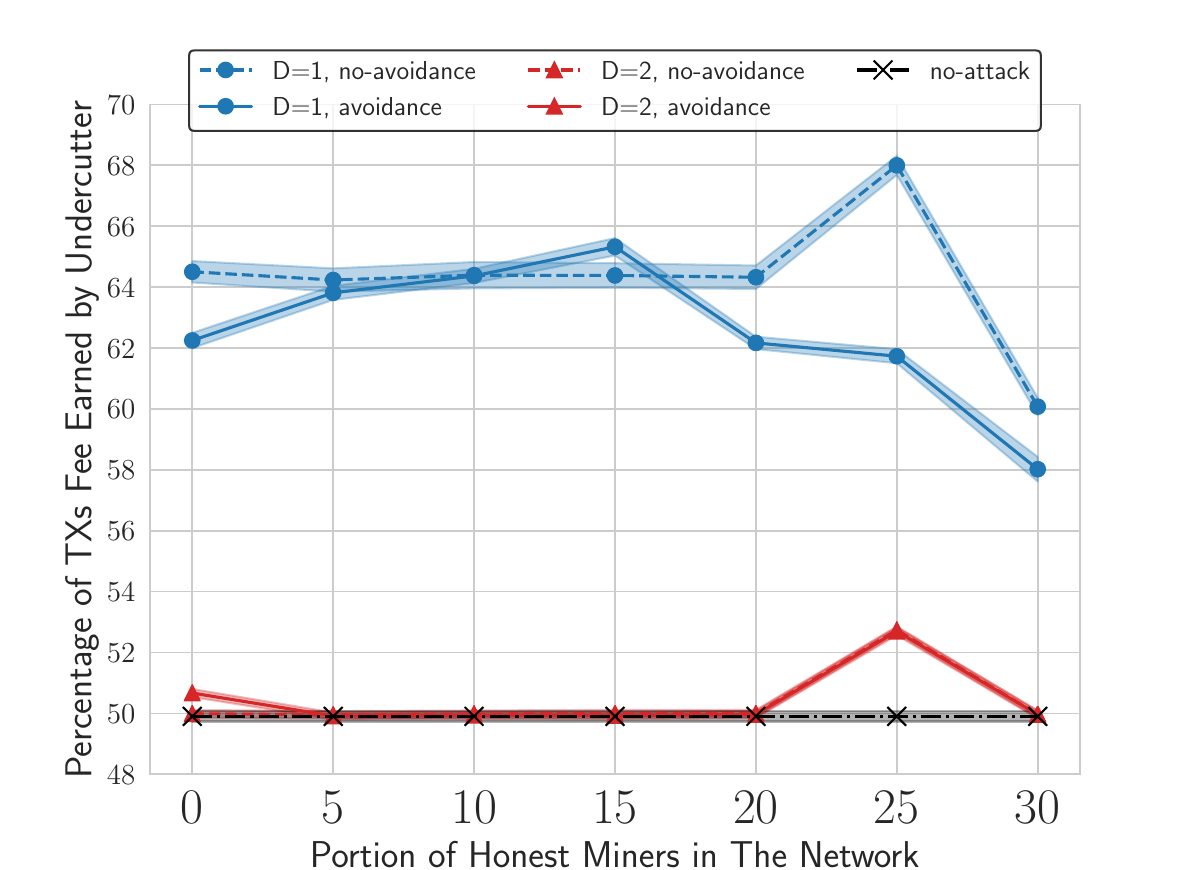}
 \caption{Artificial transactions with normal distributed fee rates: returns for $49\%$ Miner.}
 \label{fig:fakenormal49}
 \end{subfigure}
 \begin{subfigure}{0.45\textwidth}
 \centering
 \includegraphics[width=\columnwidth]{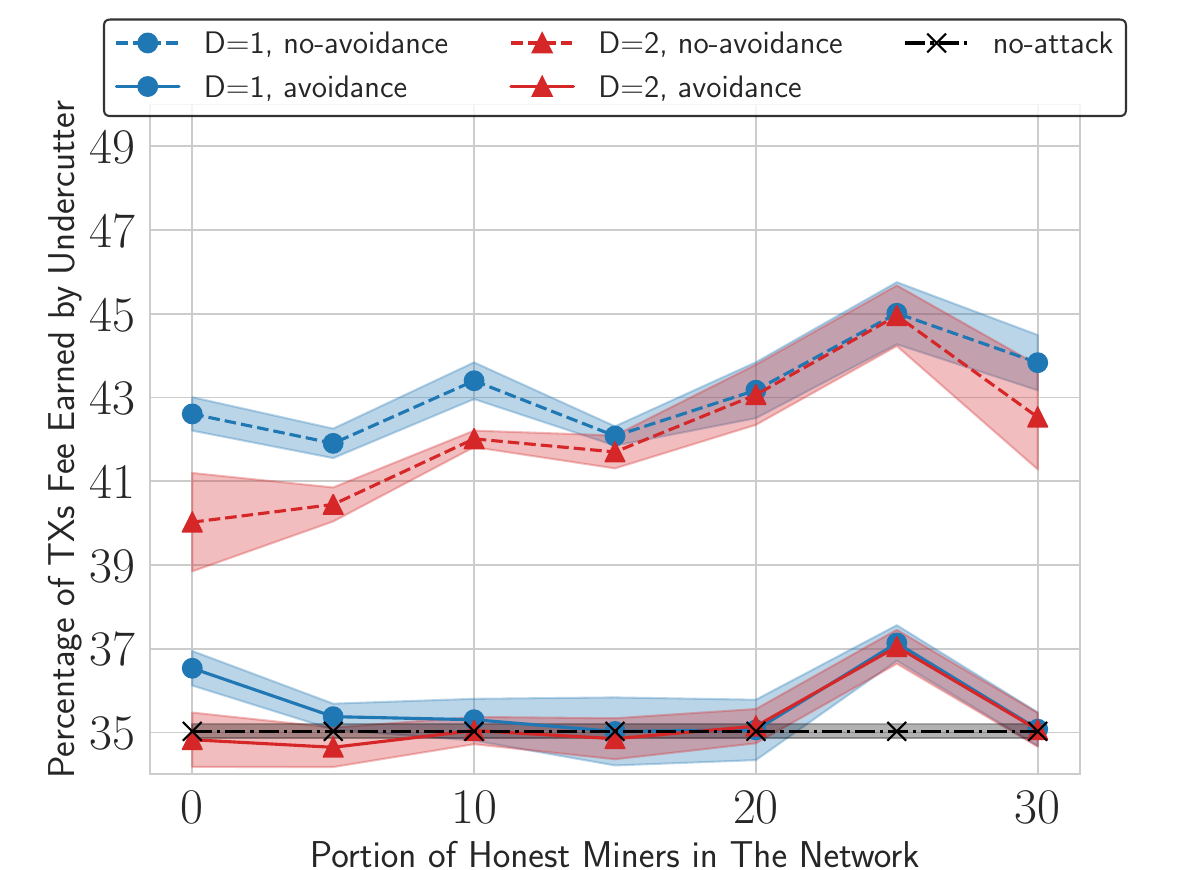}
 \caption{Monero: Returns for 35\% attacker.}
 \label{fig:monero}
 \end{subfigure}
 \caption{
 Undercutting Returns: normal runs (dashed lines) and runs with avoidance feature enabled (solid lines). 
 The shadowed band is statistics' 95\% confidence interval.
 } 
 \label{fig:bitcoin}
\end{figure*}

\subsection{Experiment Results}

\paragraph{Normal runs} Overall in a normal run, a strong undercutter can expect to earn more than fair shares by conditional undercutting as shown in figures \ref{fig:bitcoin49} and \ref{fig:monero}. \textbf{(i)} In Bitcoin runs, the 17.6\% undercutter receives on average (for $D=1$) 17.9\% shares for 0-50\% honest mining power (Figure~\ref{fig:bitcoin17}). The strong 49.9\% undercutter receives a greater profit of 60.8\% of the shares (Figure~\ref{fig:bitcoin49}). 
\textbf{(ii)} In runs with artificial transactions, the profits for $D=1,2$ bear a wider gap than with actual Bitcoin transactions (Figure~\ref{fig:fakenormal49}). 
\textbf{(iii)} In Monero runs, the 35\% undercutter obtains 43.2\% of the profit on average (for $D=1,2$) for different honest miner portions (\Cref{fig:monero}). Undercutting is especially efficient in Monero because of its small mempools, which provide limited cushion effects.

\paragraph{With undercutting avoidance} As noted by PoA, the attacker has an advantage over others in equilibrium. 
The predicted average revenue proportion (adjusted for rounds where the undercutter mines a block and attacking is unnecessary) for the 49.9\% attacker is around 63\%. \textbf{(i)} In Bitcoin actual and artificial data runs, the return proportion is close to this predicted average. Avoidance runs can result in better revenues for the undercutter if the attack \emph{cannot} be carried out to its ideal extent. That is because a large mempool along with continual incoming transactions lowers the profitability of undercutting. The implication is that if undercutting cannot be implemented ideally, avoidance can be relaxed from the exact extent. 
\textbf{(ii)} For Monero, we observe profit reduction for attackers in both margins after enabling avoidance, as shown in~\Cref{fig:monero}. 
\textbf{(iii)} Monero runs and Bitcoin runs for 17.6\% undercutter provide more straightforward results, compared to Bitcoin runs with 49.9\% attacker. Because the second undercutter in Monero has 35\% mining power, which equals the strongest undercutter's mining power and in Bitcoin, the configuration is that the second-strongest mining power is 15.3\% for 17.6\% attacker and 20\% for 49.9\% attacker. 
Overall, normal runs provide a lower bound on the profits $R_u^{lo}$ from undercutting (on the data set) and avoidance-enabled runs give an upper bound on undercutting profits $R_{a}^{up}$ (on the data set). When $R_u^lo > R_{a}^{up}$, avoidance is recommended.

\subsection{Undercutting and Avoidance in the wild}
In the real-world implementation of undercutting, system policies concerning undercutting attacks, mempool states, miner type composition, and network latency can be ever-changing. Its performance can vary across time and will be improved when one can predict future transaction arrival to better precision, e.g., via learning tools. 
Avoidance parameters can also be adjusted accordingly based on observations and learning models.

\paragraph{Latency} Network and transaction propagation latency have large impacts on mempool states, thus affecting the profitability of undercutting. The undercutter needs to advertise for the wealthy transactions left unclaimed to attract other rational miners to join an attack. 

\paragraph{Other Mitigations} 
There also exist other mitigation techniques. One such proposal is to implement the rule that transactions include the height of the latest chain head block, and they can only be included in blocks of higher height. For example, at time $t_1$, a new block is appended to the block at height $h$. Then transactions appearing during height $h$ and $h+1$ are only allowed to be included in blocks of height higher than $h$. We note that the effectiveness of this defense technique is discounted by the size of the mempool.

Another viable method is to penalize undercutters. 
To differentiate between normal forks and undercutting, one can examine the timestamp and differences of embedded transactions inside competing chain heads, and whether this happens regularly. An undercutter often starts attacking after the target block has been created and includes only part of claimable transactions. 

\paragraph{Minor changes to Bitcoin core codebase} 
We note that only light code changes in the Bitcoin core codebase are needed, which we demonstrate in this source~\cite{btccodesrc} and also describe the code snippet in Appendix~\ref{discussion}. 
\section{Conclusion}\label{sec:Conclusion}
We study the profitability of the undercutting mining strategy with the block size limit present. 
The intentional balancing of undercutting others and avoiding one's fork being undercut again demands specific conditions on the unconfirmed transaction set at the time of decision-making. Once conditions are met, an attacker can expect positive premiums. However, because such conditions are not easy to satisfy, are time-dependent (can be invalidated if new transactions arrive), and can be manipulated, it opens a door for mitigation. By applying an avoidance technique to invalidate the aforementioned conditions, miners can avoid being undercut. Avoidance encourages miners to claim fewer fees if the current bandwidth set is sufficiently wealthier than the next bandwidth set. As a result, the competition of undercutting can involuntarily promote the fair sharing of fees even in a time-variant fee system. Nevertheless, in a one-sided competition where the mining power discrepancy between the first and second strongest undercutters is large, the stronger undercutter has a natural advantage over others because it only has to defend against the weaker. 

\paragraph{Acknowledgement}
We would like to thank our shepherd Marko Vukolic and anonymous reviewers for their valuable comments. We thank Dankrad Feist for his feedback in the early stage of this project. This work has been partially supported by the National Science Foundation under grant CNS-1846316.


\bibliographystyle{IEEEtranS}
\bibliography{all}

\appendices
\section{Choice of $D$}\label{choiced}
We have looked into safe margin parameters $D=1$ and $D=2$ in the analysis. We do not explore into $D>2$ because first, from $D=1$ to $D=2$, we observe a change towards tighter attack conditions and smaller profitability. Intuitively, there are more uncertainties when safe margins increase as undercutters rely on future bandwidth set to be less wealthy than the undercutting target block. A larger safe margin potentially reduces the premiums from each attack, and attack conditions become tighter, which diminishes the total number of attacks. Second, when we apply avoidance techniques as a defense against undercutting, the avoidance for $D=1$ is the strongest. In other words, when we defend against a $D=1$ attacker, we defend against others as well. Third, the probability of catching up after being more than 2 blocks behind is small when the mining resources concentrated on the chain are not significantly large. We give more details about this argument below. 

\textbf{Problem Statement.} Let $m,n$ denotes the relative height of chain $C_0$ and chain $C_1$ after the forking point. Let $O\leq 0.5$ be the mining power on $C_1$ and $\tilde{D}=n-m$ ($|\tilde{D}|<D$). Show that for $D>2$, the probability of $C_1$ winning when $\tilde{D}\leq -2$ is small. 

\begin{proof}
Let $X, Y$ be the random variables for the waiting time between two blocks on $C_0$ and $C_1$. 
We compute the probability of $C_1$ winning as follows: 
\begin{align*}
 p = \Pr[\text{$C_1$ wins}]=\sum_{i=0}^{\infty}\Pr[(D-\tilde{D}+i)\cdot Y<(i+1)\cdot X] 
\end{align*}

We know that $D-\tilde{D}\geq 5$.
\begin{align*}
 p &=\sum_{i=0}^{\infty} O^{D-\tilde{D}+i}(1-O)^i\leq \sum_{i=0}^{\infty} O^{5+i}(1-O)^i
\end{align*}

We know $O^5$ and $(O (1-O))^i$ take maximum at $O=0.5$. Let $O=0.5$, we have
\begin{align*}
 p <\sum_{i=0}^{\infty} (\frac{1}{2})^{5+2i}=\frac{\sum_{i=0}^{\infty}4^{-i}}{32}=\frac{1}{32}\lim_{n\rightarrow \infty} \frac{1-4^{-n}}{1-1/4}=\frac{1}{24}
\end{align*}
We consider this probability to be relatively small. 
Similarly, we can compute for $O=0.6$ that $p<0.10$, for $O=0.7$ that $p<0.21$, for $O=0.8$ that $p<0.39$ and for $O=0.9$ that $p<0.65$. Therefore, when $O$ is not significantly large, the probability of $C_1$ winning when $\tilde{D}\leq -2$ is small.
\end{proof}

\section{Changes to Bitcoin Core codebase}\label{discussion}
The changes are mainly contained in ``$miner.cpp$'' with the parameter for threshold value $T$ being set in consensus file ``$consensus.h$'' and ``$policy.h$''. Major changes specifically reside in $addPackageTxs()$ and a utility function $RemoveFromBlock()$ is added. We show the code snippet we add to the transaction selection function as follows:

\onecolumn
\begin{lstlisting}[language=C++, title=Changes to function BlockAssembler::addPackageTxs]
// This transaction selection algorithm orders the mempool based
// on feerate of a transaction including all unconfirmed ancestors.
// Since we don't remove transactions from the mempool as we select them
// for block inclusion, we need an alternate method of updating feerate
// of a transaction with its not-yet-selected ancestors as we go.
// This is accomplished by walking the in-mempool descendants of selected
// transactions and storing a temporary modified state in mapModifiedTxs.
// Each time through the loop, we compare the best transaction in
// mapModifiedTxs with the next transaction in the mempool to decide what
// transaction package to work on next.
void BlockAssembler::addPackageTxs(int &nPackagesSelected, int &nDescendantsUpdated)
{
    // until line 331
    while (mi != m_mempool.mapTx.get<ancestor_score>().end() || !mapModifiedTx.empty()) {
        // ... until line 395
        if (!TestPackage(packageSize, packageSigOpsCost)) {
            // ... until line 406
            if (TestPackageForBS(packageSize, packageSigOpsCost)) {
                // add the failed transaction to next bandwidth set
                CTxMemPool::setEntries ancestors;
                uint64_t nNoLimit = std::numeric_limits<uint64_t>::max();
                std::string dummy;
                m_mempool.CalculateMemPoolAncestors(*iter, ancestors, nNoLimit, nNoLimit, nNoLimit, nNoLimit, dummy, false);

                onlyUnconfirmed(ancestors);
                ancestors.insert(iter);

                // Test if all tx's are Final
                if (TestPackageTransactions(ancestors)) {
                    // Package can be added. Sort the entries in a valid order.
                    std::vector<CTxMemPool::txiter> sortedEntries;
                    SortForBlock(ancestors, sortedEntries);
                    for (size_t i=0; i<sortedEntries.size(); ++i) {
                        nFeesNext += sortedEntries[i]->GetFee();
                        nNextBlockWeight += sortedEntries[i]->GetTxWeight();
                        nNextBlockSigOpsCost += sortedEntries[i]->GetSigOpCost();
                    }
                }
            }

            if (nConsecutiveFailed > MAX_CONSECUTIVE_FAILURES && nBlockWeight >
                    nBlockMaxWeight - 4000 && nNextBlockWeight >
                    nBlockMaxWeight - 10000 ) {
                // Give up if we're close to full and haven't succeeded in a while
                assert(!nFees);
                float gamma = nFeesNext/nFees;
                float T = 0.63; // as an example, 30% strongest rational miner min{0.5,0.63}
                // T needs to be in consensus file
                CAmount nFeesCurrent = 0;
                if (gamma < T) {
                    nFeesCurrent = std::floor(nFees* (1+gamma) / (1+T));
                    // iterate over transactions at the end of the current block
                    CTxMemPool::setEntries::iterator iit = inBlock.end(); 
                    // CTxMemPool::txiter iit = inBlock.end(); 
                    while (nFeesCurrent < nFees) {
                        failedTx.insert(*iit);
                        RemoveFromBlock(*iit);
                        iit--;
                    }
                }
                break;
            }
            continue;
        }
    // until the end
    }
}

void BlockAssembler::RemoveFromBlock(CTxMemPool::txiter iter)
{
    pblocktemplate->block.vtx.pop_back();
    pblocktemplate->vTxFees.pop_back();
    pblocktemplate->vTxSigOpsCost.pop_back();
    nBlockWeight -= iter->GetTxWeight();
    --nBlockTx;
    nBlockSigOpsCost -= iter->GetSigOpCost();
    nFees -= iter->GetFee();
    inBlock.erase(iter);

    bool fPrintPriority = gArgs.GetBoolArg("-printpriority", DEFAULT_PRINTPRIORITY);
    if (fPrintPriority) {
        LogPrintf("fee %s txid %s\n",
                    CFeeRate(iter->GetModifiedFee(), iter->GetTxSize()).ToString(),
                    iter->GetTx().GetHash().ToString());
    }
}

// in miner.h : class BlockAssembler
    uint64_t nNextBlockWeight;
    uint64_t nNextBlockSigOpsCost;
    CAmount nFeesNext = 0;
\end{lstlisting}

\end{document}